\documentclass[lettersize,journal]{IEEEtran}
\usepackage{amsmath,amsfonts}
\usepackage{algorithmic}
\usepackage{algorithm}
\usepackage{array}
\usepackage[caption=false,font=normalsize,labelfont=sf,textfont=sf]{subfig}
\usepackage{textcomp}
\usepackage{url}
\usepackage{verbatim}
\usepackage{graphicx}
\usepackage{cite}
\usepackage{doi}
\usepackage{xcolor}
\usepackage[utf8]{inputenc}
\usepackage[T1]{fontenc}
\usepackage{etoolbox}
\newcommand{\hide}[1]{}

\makeatletter
\g@addto@macro\normalsize{%
  \setlength{\abovedisplayskip}{1.2ex plus 2pt minus 1pt}%
  \setlength{\belowdisplayskip}{1.2ex plus 2pt minus 1pt}%
  \setlength{\abovedisplayshortskip}{0pt plus 1pt}%
  \setlength{\belowdisplayshortskip}{0.8ex plus 2pt minus 1pt}%
}
\makeatother

\begin{document}
\bstctlcite{IEEEexample:BSTcontrol}

\title{Link Both Views: Cross-Image Routing Attention for Dual-Endpoint V2I Beam-Pair Power Prediction}

\author{Shiliang Lu,~\IEEEmembership{Graduate~Student~Member,~IEEE,} Lu Bai,~\IEEEmembership{Senior Member,~IEEE,} Ziwei Huang,~\IEEEmembership{Member,~IEEE,} and Xiang Cheng,~\IEEEmembership{Fellow,~IEEE}
        % <-this % stops a space
\thanks{
Shiliang Lu and Lu Bai are with the Joint SDU-NTU Centre for Artificial Intelligence Research (C-FAIR), Shandong University, Jinan 250101, China. Shiliang Lu is also with the School of Software, Shandong University (e-mail: shiliang.lu@mail.sdu.edu.cn; lubai@sdu.edu.cn).
}
\thanks{
Ziwei Huang and Xiang Cheng are with the State Key Laboratory of Photonics and Communications, School of Electronics, Peking University, Beijing 100871, China (e-mail: ziweihuang@pku.edu.cn; xiangcheng@pku.edu.cn).
}
}

\markboth{IEEE Wireless Communications Letters,~Vol.~XX, No.~XX, XX~2026}%
{Shell \MakeLowercase{\textit{et al.}}: A Sample Article Using IEEEtran.cls for IEEE Journals}


\maketitle 

\begin{abstract}
For sensing-aided communications, single-endpoint sensing offers only a partial view of propagation, motivating link-centric sensing across both endpoints. We propose cross-image routing attention (CIRA) for sparse association and mutual enhancement of propagation-relevant features from dual-endpoint RGB images. Building on CIRA, CiFormer predicts millimeter-wave (mmWave) beam-pair power matrices for vehicle-to-infrastructure (V2I) communications. Experiments show that CiFormer reduces mean absolute error and improves top-1 beam-pair selection accuracy by 3.15 percentage points over the strongest baseline. Visualizations and ablation studies support dual-endpoint sensing and selective cross-view interaction in capturing propagation-relevant dependencies, suggesting broader applications in wireless tasks beyond beam prediction.
\end{abstract}


\section{Introduction}

\IEEEPARstart{A}{rtificial intelligence} (AI) and machine learning (ML) are revolutionizing wireless network design, enabling more efficient, adaptive, and autonomous operation \cite{cheng2025integrated}. In the Internet of Vehicles (IoV), environment sensing further supports this evolution by providing observations of the propagation environment, offering new opportunities for informed and proactive decision-making \cite{\hide{cheng2025integrated,}bai2026mmicm}. However, leveraging high-dimensional environment observations for IoV communication system optimization is far from straightforward. This difficulty arises from the complex and dynamic relationship between environment observations and wireless propagation. Consequently, developing effective mechanisms to identify, represent, and exploit propagation-relevant sensing information remains a key challenge for environment-aware intelligent communications.

Most existing sensing-aided approaches rely on single-endpoint sensing, with sensors deployed at either the vehicle or the infrastructure endpoint \cite{charan2025camera,xu2023computer}. However, a single sensing endpoint captures only a partial view of the propagation environment due to its limited sensing coverage \cite{\hide{bai2026mmicm, }lin2024multicamera}. This limitation reveals a structural mismatch: a wireless link is inherently defined by two communication endpoints, i.e., the transmitter (Tx) and receiver (Rx), whereas single-endpoint sensing observes the propagation environment from only one side. Accordingly, this mismatch highlights the need to better align sensing with the two-endpoint structure of the wireless link, motivating a transition from endpoint-local sensing to link-centric sensing.

Nevertheless, effective characterization of the wireless link requires more than simply acquiring dual-endpoint observations. The two endpoint views are captured from distinct positions and orientations, resulting in substantial differences in perspective and semantic content. Given these differences, independently processing the two views or directly concatenating their features is insufficient to capture the link-relevant cross-view relationships \cite{zeng2026bev}. Therefore, dual-endpoint sensing calls for an effective mechanism to identify and associate propagation-relevant information across the two views, jointly highlighting regions in both views associated with propagation paths that couple Tx-side departure and Rx-side arrival, rather than treating them as two independent observations. More broadly, capturing propagation-relevant relationships across the Tx and Rx views could support a range of tasks, including channel modeling, beam management, and resource allocation.

A representative link-centric task arises in millimeter-wave (mmWave) vehicle-to-infrastructure (V2I) communications. High mobility, dynamic blockage, and scattering can induce rapid channel fluctuations and consequently change the received powers of different beam pairs \cite{tan2024beamalignment,wang2018situational}. A beam-pair power matrix (BPPM) characterizes the received powers across all candidate beam pairs, providing a more comprehensive description of the beam-domain link state than the strongest beam-pair index alone \cite{wang2018situational}. Accordingly, BPPM prediction depends jointly on the states of both endpoints and the propagation-relevant information of their surroundings. However, existing sensing-aided beam prediction approaches \cite{charan2025camera,xu2023computer,lin2024multicamera} predominantly exploit observations from only one side of the communication link, leaving link-relevant information across the two endpoint views insufficiently exploited.

To address these challenges, we develop CiFormer, a novel link-centric framework that captures propagation-relevant information through interactions between Tx- and Rx-side images. Then, this resulting joint representation is used for BPPM prediction. The main contributions and novelties are summarized as follows:
\begin{itemize}
    \item A novel cross-image routing attention (CIRA) mechanism is proposed. Specifically designed for sensing-aided wireless  communications, CIRA recasts visual attention from individual views to dual-endpoint views through a propagation-inspired coupling mechanism, making it the first attention mechanism to incorporate link-level dependencies into visual perception. Motivated by endpoint co-visibility, it links both views through sparse cross-view associations and potentially supports broader wireless environment-aware communication tasks.

    \item A link-centric BPPM prediction framework, CiFormer, is proposed. Through CIRA, Tx- and Rx-side visual features are enabled to interact and form mutually enhanced representations. The resulting joint propagation representation alleviates single-endpoint partial observability, improving power prediction across candidate beam pairs and enabling more reliable beam-pair selection.

\end{itemize}

\section{System Model and Problem Formulation}
Consider a V2I link from a vehicle Tx with $N_{\mathrm{Tx}}$ antennas to a roadside base station (BS) Rx with $N_{\mathrm{Rx}}$ antennas. The corresponding channel is denoted by $\mathbf{H}\in\mathbb{C}^{N_{\mathrm{Rx}}\times N_{\mathrm{Tx}}}$. At each link snapshot, an RGB image $\mathbf{I}^{v}$ is acquired at each viewpoint $v\in\mathcal{V}=\{\mathrm{Tx},\mathrm{Rx}\}$, where $\bar{v}$ denotes the opposite side of $v$. The paired observations $\left(\mathbf{I}^{v},\mathbf{I}^{\bar{v}}\right)$ provide complementary views of the propagation environment.

To characterize the link in the beam domain, fixed discrete Fourier transform (DFT)-based codebooks with unit-norm codewords are employed at both endpoints~\cite{yang2010dft}. Let $\mathbf{F}=[\mathbf{f}_0,\ldots,\mathbf{f}_{P-1}]\in\mathbb{C}^{N_{\mathrm{Tx}}\times P}$ and $\mathbf{W}=[\mathbf{w}_0,\ldots,\mathbf{w}_{Q-1}]\in\mathbb{C}^{N_{\mathrm{Rx}}\times Q}$ denote the Tx precoding and Rx combining codebooks, respectively, where $P$ and $Q$ denote the numbers of Tx and Rx beams. The codebook beams cover different azimuth directions, with elevation steering fixed at broadside. Following~\cite{wang2018situational}, each beam pair has a power gain $G_{q,p}=\left|\mathbf{w}_q^{H}\mathbf{H}\mathbf{f}_p\right|^2$, and the complete BPPM is computed in matrix form as
\begin{equation}
\mathbf{G} = \left|\mathbf{W}^{H}\mathbf{H}\mathbf{F}\right|^{\odot 2}.
\label{eq:bppm}
\end{equation}
Here, $(\cdot)^H$ denotes the conjugate transpose, $|\cdot|^{\odot 2}$ denotes the entrywise squared magnitude, and $\mathbf{G}\in\mathbb{R}_{\geq 0}^{Q\times P}$ contains the dimensionless channel power gains for all $QP$ candidate beam pairs. Its row index $q\in\{0,\ldots,Q-1\}$ and column index $p\in\{0,\ldots,P-1\}$ correspond to the Rx and Tx beams, respectively. For a fixed transmit power, these gains are proportional to the received signal powers.

Given the endpoint-indexed image pair, the prediction target, i.e., the entire BPPM, is expressed in dB as $Y_{q,p}=10\log_{10}(\max\{G_{q,p},\epsilon\})$, where $\epsilon>0$ ensures numerical stability, and the target matrix is $\mathbf{Y}=[Y_{q,p}]\in\mathbb{R}^{Q\times P}$. The prediction problem is formulated as learning the mapping
\begin{equation}
\widehat{\mathbf{Y}}
=f_{\theta}\!\left(\{\mathbf{I}^{v}\}_{v\in\mathcal{V}}\right)
\in\mathbb{R}^{Q\times P}
\label{eq:visual_bppm_prediction}
\end{equation}
where $\theta$ denotes the learnable parameters and the input images retain their endpoint identities. Training uses paired endpoint images and their corresponding BPPM labels. $\mathbf{H}$ is used only to construct the labels and is not an input to the predictor. The predicted beam pair is selected as $(\hat{q},\hat{p})=\arg\max_{q,p}\widehat{Y}_{q,p}$~\cite{wang2018situational}.

Overall, this formulation connects dual-endpoint visual observations with the power gains of all candidate beam pairs. The following section presents CiFormer for learning this mapping through propagation-relevant cross-view interactions.

\section{Proposed Method}
In this letter, we propose CiFormer, which couples propagation-relevant cross-view feature association with structured beam-pair power prediction. CIRA provides the underlying interaction mechanism, while the BPPM prediction head estimates beam-pair powers from the jointly enhanced endpoint features.

\subsection{Cross-Image Routing Attention}

\begin{figure}[!t]
	\centering	\includegraphics[width=0.99\columnwidth]{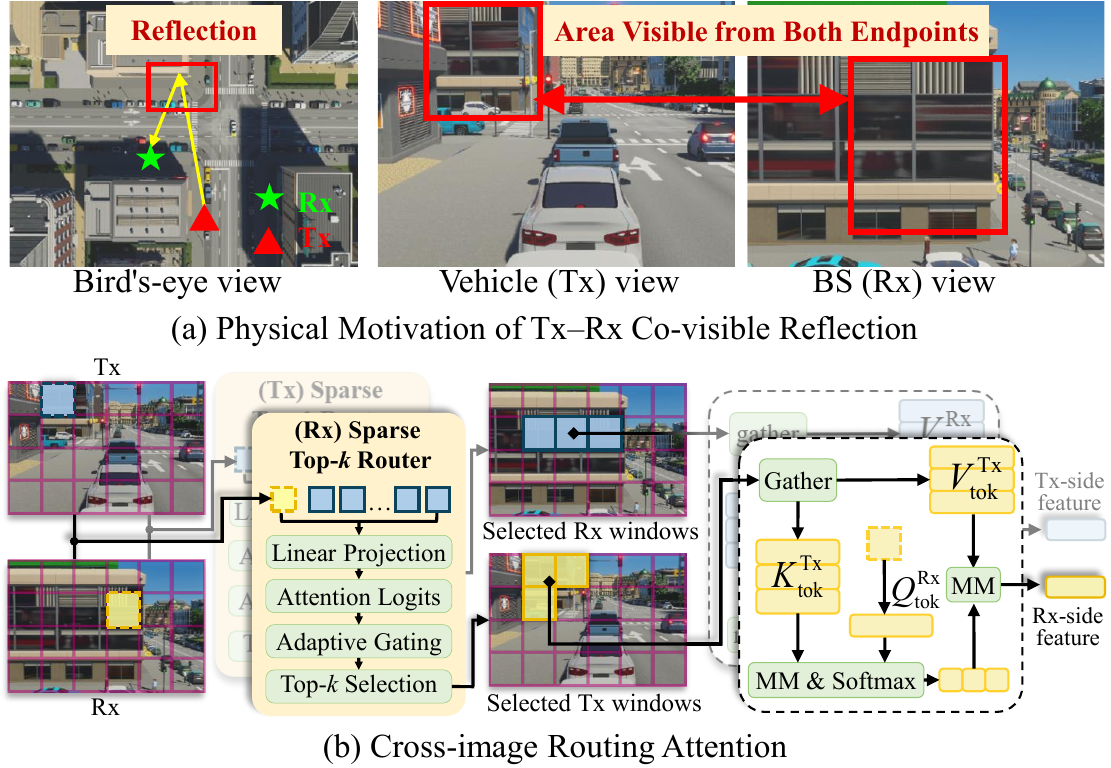}
	\caption{Illustration of the proposed cross-image routing attention. (a) Physical motivation of Tx--Rx co-visible reflection in wireless channels, where objects simultaneously visible from both views are more likely to contain scatterers that give rise to NLoS paths. (b) Cross-image routing attention, which enhances the modeling of potential reflection-related windows guided by the Tx--Rx co-visible propagation mechanism.}
	\label{CIRA}
\end{figure}

\begin{figure*}[!t]
	\centering	\includegraphics[width=1.0\textwidth]{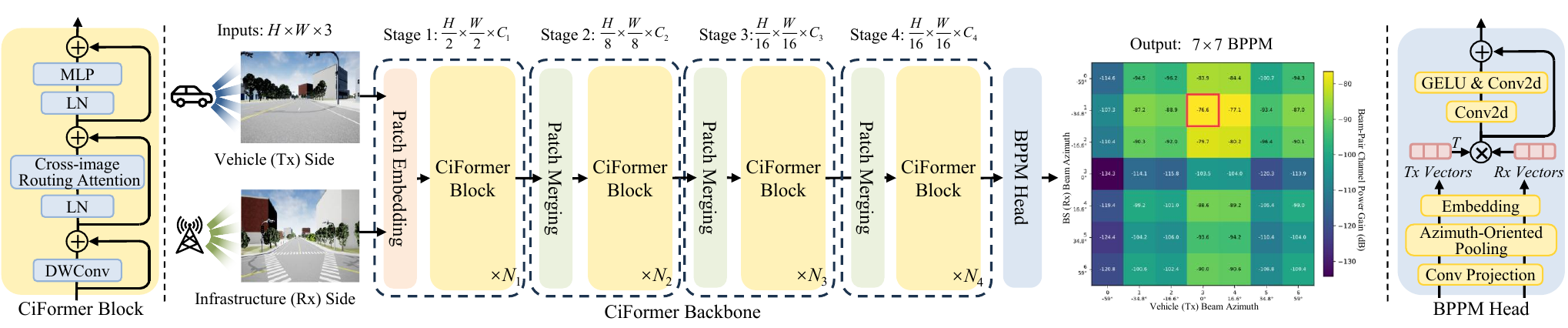}
	\caption{Overall architecture of CiFormer, with detailed structures of the CiFormer block on the left and the BPPM prediction head on the right.}
	\label{CiFormer}
\end{figure*}

The RGB images captured at both link ends provide visual semantics of the propagation environment. However, the Tx and Rx cameras observe the same physical scene from substantially different viewpoints, resulting in disparities in object appearance, scale, and occlusion. This challenge is further motivated by wireless channel propagation, where a single-bounce non-line-of-sight (NLoS) path requires a scattering region to be visible from both Tx and Rx, as illustrated in Fig.~\ref{CIRA}(a). Therefore, co-visible regions are more informative for characterizing such propagation, whereas dense attention computes interactions for all regions without accounting for their propagation relevance. Motivated by this observation, we propose CIRA, which establishes sparse cross-view interactions through a learnable top-$k$ routing mechanism. As shown in Fig.~\ref{CIRA}(b), CIRA extends BiFormer's bi-level routing~\cite{zhu2023biformer} to cross-image attention, where queries from one view are routed to the most relevant windows in the other view rather than within the same feature map. By focusing cross-view interaction on selected windows, CIRA reduces redundant computation while enhancing propagation-relevant feature association. Algorithm~\ref{alg:cira} summarizes the overall procedure.

% Directly concatenating or globally attending across the two views therefore introduces redundant computation while limiting to effectively capture cross-view correspondence.

\par\vspace{0.4em}
\noindent\textbf{Window partitioning and QKV projection.}
Given RGB features from two views $\mathbf{X}^v \in \mathbb{R}^{H \times W \times C}$, we first partition each into $N = n_h \times n_w$ windows of size $s_h \times s_w$, yielding $\mathbf{X}^{v}_{p}\in\mathbb{R}^{s_hs_w\times N\times C}$. Each window is then mapped to query, key, and value via learnable linear projection as
\begin{equation}
    \mathbf{Q}^{v}_{\text{tok}} =  \mathbf{X}^{v}_{p} \mathbf{W}^q,\quad
    \mathbf{K}^{v}_{\text{tok}} =  \mathbf{X}^{v}_{p} \mathbf{W}^k,\quad
    \mathbf{V}^{v}_{\text{tok}} =  \mathbf{X}^{v}_{p} \mathbf{W}^v.
\end{equation}
Note that $\mathbf{W}^q, \mathbf{W}^k, \mathbf{W}^v$ are shared across both views to ensure a consistent embedding space. 
To reduce the cost of subsequent attention, the key and value features are spatially downsampled via adaptive average pooling to $k_v \!\times\! k_v$ tokens, with $k_v=4$ following \cite{zhu2023biformer}, yielding $\mathbf{K}^{v}_{\text{tok}}, \mathbf{V}^{v}_{\text{tok}} \in \mathbb{R}^{k_v^2 \times N \times C}$.

\par\vspace{0.4em}
\noindent\textbf{Window-to-window cross-image routing.}
To capture the co-visible prior, we perform routing at the window level. Specifically, each window is summarized by average pooling over the spatial dimension to obtain window-level descriptors $\mathbf{Q}^{v}_{\text{win}}, \mathbf{K}^{v}_{\text{win}} \in \mathbb{R}^{N \times C}$. A learnable projection $\mathbf{W}^{r}\in\mathbb{R}^{C\times C}$ is adopted to enhance their discriminability, after which the cross-view affinity is computed as
% after which the scaled dot-product attention logits form an adjacency matrix of window-level affinity between Tx and Rx, which can be written as
\begin{equation}
    \mathbf{A}^{v} =\frac{(\mathbf{Q}^{v}_{\text{win}} \mathbf{W}^{r})(\mathbf{K}^{\bar v}_{\text{win}} \mathbf{W}^{r})^\top
    }{\sqrt{C}}\in \mathbb{R}^{N \times N}
\end{equation}
where $\mathbf{A}^{v}$, i.e., affinity score, quantifies the semantic similarity between two windows. For each window in view $v$, the top-$k$ most relevant windows from view $\bar v$ are retained, yielding a routing index set $\mathcal{I}^{v} = \text{topkIndex}(\mathbf{A}^{v})$ and routing weights $r^{v}_{ij}$ via softmax, where $i$ and $j$ denote the query and routed key window indices, respectively. Then, a soft query gate $g(\cdot)$ modulates routing weights to suppress unreliable queries.

\begin{algorithm}[!t]
\footnotesize
\caption{Cross-Image Routing Attention}
\label{alg:cira}
\textbf{Input:} Tx, Rx features $\mathbf{X}^{\text{Tx}},\mathbf{X}^{\text{Rx}}\in\mathbb{R}^{H\times W\times C}$ \\
\textbf{Output:} Tx, Rx features $\mathbf{X}^{\text{Tx}}_{\text{RGB}},\mathbf{X}^{\text{Rx}}_{\text{RGB}}$\\[-3ex]
\begin{algorithmic}[1]
\STATE Partition $\mathbf{X}^{\text{Tx}},\mathbf{X}^{\text{Rx}}$ into $N$ windows%: $\mathbf{X}^{\text{Tx}}_{p},\mathbf{X}^{\text{Rx}}_{p}$
\STATE Shared QKV projection; downsample $\mathbf{K},\mathbf{V}$ to $k_v \times k_v$
\STATE Obtain $\mathbf{Q}_{\text{win}},\mathbf{K}_{\text{win}}$ by average pooling

\FOR{$v\in\mathcal{V}=\{\mathrm{Tx},\mathrm{Rx}\}$}

\STATE $\mathbf{A} = (\mathbf{Q}^{v}_{\text{win}}\mathbf{W}^r)(\mathbf{K}^{\bar v}_{\text{win}}\mathbf{W}^r)^\top / \sqrt{C}$
\STATE $\mathcal{I} = \mathrm{TopKIndex}(\mathbf{A},k)$
\STATE $\mathbf{r} = \mathrm{Softmax}(\mathrm{TopK}(\mathbf{A},k))$
\STATE $\mathbf{r} \leftarrow \mathbf{g}(\mathbf{Q}^{v}_{\text{win}}) {\odot} \mathbf{r}$ 

\FOR{$i=1$ to $N$}
    \STATE $(\mathbf{K}^{\bar v}_i,\mathbf{V}^{\bar v}_i)=\mathrm{Gather}(\mathbf{K}^{\bar v},\mathbf{V}^{\bar v},\mathcal{I}_i,\mathbf{r}_i)$
    \STATE $\mathbf{B}_i=\mathrm{MLP}(\phi_q(\mathbf{Q}^{v}_{\text{win},i})-\phi_k(\mathbf{K}^{\bar v}_{\text{win},\mathcal{I}_i}))$
    \STATE $\mathbf{O}^{v}_i=\mathrm{Softmax}(\mathbf{Q}^{v}_{i}(\mathbf{K}^{\bar v}_i)^\top/\sqrt{C}+\mathbf{B}_i)\mathbf{V}^{\bar v}_i$
\ENDFOR

\STATE $\mathbf{X}^{v}_{\text{RGB}}=\mathbf{W}^o(\mathrm{Unpatchify}(\{\mathbf{O}^{v}_i\})+\mathrm{LCE}(\mathbf{V}^{v}))$
\ENDFOR
\STATE \textbf{Return} $\mathbf{X}^{\text{Tx}}_{\text{RGB}},\mathbf{X}^{\text{Rx}}_{\text{RGB}}$
\end{algorithmic}
\end{algorithm}

\par\vspace{0.4em}
\noindent\textbf{Semantic-aware bias.}
Without a shared geometric reference, conventional relative position biases offer limited benefit. Instead, we introduce a semantic-aware bias (SAB) that derives a pairwise additive term from the feature discrepancies between a query window and its routed key windows, formulated as
\begin{equation}
    \mathbf{B}^{v}_i = \text{MLP}(\phi_q(\mathbf{Q}^{v}_{\text{win},i}) - \phi_k(\mathbf{K}^{\bar v}_{\text{win},\mathcal{I}^{v}_i}))
\end{equation}
where $\phi_q, \phi_k$ are learnable linear projections, MLP denotes a multilayer perceptron, and $\mathbf{B}^{v}_{i}$ provides the additive bias for the top-$k$ routed windows. 

\par\vspace{0.4em}
\noindent\textbf{Token-to-token cross-image attention.}
With the routing index set $\mathcal{I}^{v}$ and weights $r^{v}_{ij}$ from window-level routing, the downsampled KV tokens of the target window are gathered for efficient computation \cite{zhu2023biformer}. For the $i$-th Tx window, $\mathbf{K}^{\bar v}_i,\mathbf{V}^{\bar v}_i$ can be given as
\begin{equation}
    \mathbf{K}^{\bar v}_i =
    \text{gather}(\mathbf{K}^{\bar v}_{\text{tok}}, \mathcal{I}^{v}_i, r^{v}_{i}), \quad
    \mathbf{V}^{\bar v}_i =
    \text{gather}(\mathbf{V}^{\bar v}_{\text{tok}}, \mathcal{I}^{v}_i, r^{v}_{i}).
\end{equation}
%where $\mathcal{I}^{v}_i$ indexes the top-$k$ routed windows of $v$, and 
Then, the gathered tokens are re-weighted by the routing scores $r^{v}_{i}$ before being stacked as $\mathbf{K}^{\bar v}_i, \mathbf{V}^{\bar v}_i \!\in\! \mathbb{R}^{k k_v^2 \times C}$. The token-level cross-attention, augmented by the semantic-aware bias $\mathbf{B}^{v}_i$, is written as
\begin{equation}
    \mathbf{O}^{v}_i = \text{Softmax}\Bigl(\frac{\mathbf{Q}^{v}_{\text{tok},i} (\mathbf{K}^{\bar v}_i)^\top}{\sqrt{C}}
    + \mathbf{B}^{v}_i\Bigr)\mathbf{V}^{\bar v}_i
\end{equation}
where $\mathbf{Q}^{v}_{\text{tok},i}\in\mathbb{R}^{s_hs_w\times C}$ contains the query tokens in window $i$ of view $v$.
% $\mathbf{B}^{\text{x}}_i$ is broadcast from the window level to all intra-window token pairs. 
The reverse direction is computed symmetrically.
Window outputs are merged into the spatial grid and augmented by local context enhancement $\text{LCE}(\cdot)$ following~\cite{zhu2023biformer}, a $3\times3$ depthwise convolution over $\mathbf{V}^{v}_{\text{tok}}$. Finally, a projection $\mathbf{W}^o \!\in\! \mathbb{R}^{C \times C}$ produces the fused output as
\begin{equation}
    \mathbf{X}^{v}_{\text{RGB}} = \mathbf{W}^o (\text{Unpatchify}(\{\mathbf{O}^{v}_i\}_{i=1}^{N})+ \text{LCE}(\mathbf{V}^{v}_{\text{tok}})).
\end{equation}

\subsection{CiFormer Architecture and BPPM Prediction}
Building on CIRA, CiFormer realizes~\eqref{eq:visual_bppm_prediction} through a four-stage backbone and an azimuth-aware head that explicitly models beam-pair interactions.

\par\vspace{0.4em}
\noindent\textbf{Backbone architecture.}
As illustrated in Fig.~\ref{CiFormer}, the backbone combines convolutional downsampling with stacked CIRA blocks for cross-view interaction. Each CIRA-based block comprises a $3\times3$ depthwise convolution for implicit positional encoding, CIRA, and a two-layer MLP. Final projection convolutions produce the endpoint features $\mathbf{X}_{\mathrm{out}}^v\in\mathbb{R}^{H_f\times W_f\times C_f}$, which retain spatial information for subsequent prediction.

\par\vspace{0.4em}
\noindent\textbf{BPPM prediction head.}
As illustrated in Fig.~\ref{CiFormer}, the head first applies convolutional projection followed by azimuth-oriented pooling. Since the codebooks scan azimuth with fixed elevation, pooling aggregates features over height while preserving horizontal order. Specifically, the pooled feature at column $w$ is computed as
\begin{equation}
\mathbf{s}_w^v
=\frac{1-\beta^v}{H_f}\sum_{h=1}^{H_f}\mathbf{z}_{h,w}^v
+\beta^v\sum_{h=1}^{H_f}\alpha_{h,w}^v\mathbf{z}_{h,w}^v
\label{eq:height_aggregation}
\end{equation}
where $\mathbf{z}_{h,w}^v$ is the projected feature, $\alpha_{h,w}^v$ is a learned attention weight normalized over height, and $\beta^v\in(0,1)$ controls the mixture. 
%This combines full-height context with emphasis on informative regions while retaining horizontal structure for beam embedding. 
The pooled features are then embedded into a vector $\mathbf{e}_m^v$ for each codebook beam $m$ at view $v$.

To construct the BPPM, the head computes dot products between the two sets of beam vectors. Taking $v=\mathrm{Tx}$ and following the BPPM row--column convention, the score for Rx beam $q$ and Tx beam $p$ is
\begin{equation}
Y^{\mathrm{dot}}_{q,p} =\frac{(\mathbf{e}^{\mathrm{\bar{v}}}_q)^\top
\mathbf{e}^{\mathrm{v}}_p}{\sqrt{D}}
\end{equation}
where $D$ denotes the beam embedding dimension. Collecting all scores forms the BPPM $\mathbf{Y}^{\mathrm{dot}}\in\mathbb{R}^{Q\times P}$. The resulting BPPM is refined by a residual branch comprising two $3\times3$ convolutions with an intermediate GELU. CiFormer is trained end to end on standardized absolute-dB BPPM targets using a weighted Huber loss as
\begin{equation}
\mathcal{L}=\frac{1}{QP}\sum_{q=0}^{Q-1}\sum_{p=0}^{P-1}w_{q,p}\, \rho_{\delta}\!\left(\widehat{Y}_{q,p}-Y_{q,p}\right)
\label{eq:bppm_loss}
\end{equation}
where $\rho_{\delta}$ is the Huber-type loss and $w_{q,p}$ is a mean-normalized weight derived from the ground-truth , thereby prioritizing the optimal beam pair during training.

Overall, CIRA provides a general mechanism for propagation-relevant cross-view interaction through reciprocal feature selection and enhancement. Building on this, CiFormer integrates CIRA with a dedicated BPPM prediction head to jointly exploit dual-endpoint observations and realize the BPPM prediction mapping in~\eqref{eq:visual_bppm_prediction}, as formulated in Section~II.

\section{Experimental Results}
\subsection{Simulation Setup}
\begin{figure}[!t]
	\centering	\includegraphics[width=0.99\columnwidth]{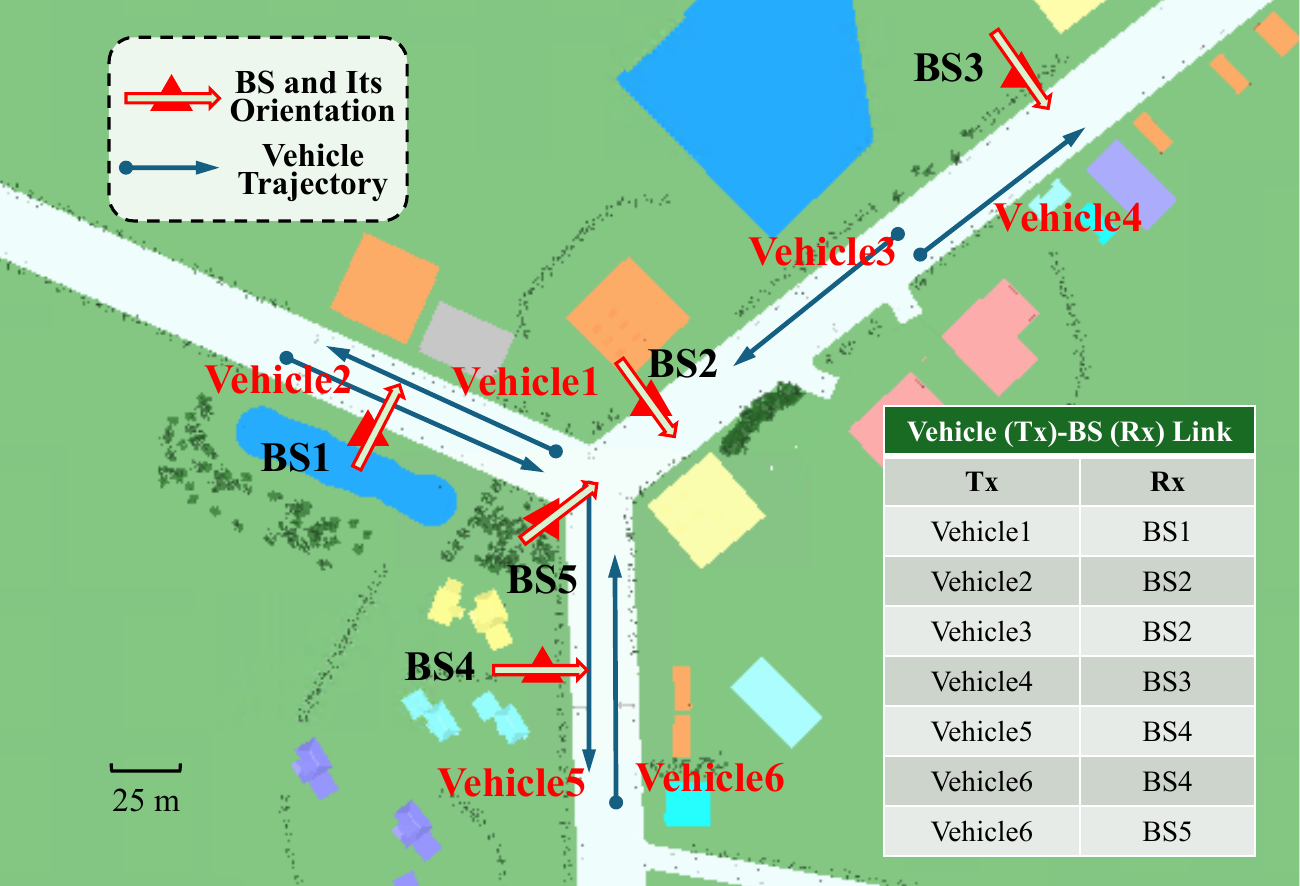}
	\caption{Bird's-eye view of the V2I simulation scenario, with endpoint locations, vehicle trajectories, BS orientations, and seven vehicle (Tx)--BS (Rx) links.}
	\label{fig:scenario}
\end{figure}

Following the aligned sensing--communication data-generation methodology in~\cite{cheng2023m3sc}, we construct the Dual-endpoint Sensing Dataset for Beam-pair Power Prediction (DuSen-BP$^3$). As illustrated in Fig.~\ref{fig:scenario}, DuSen-BP$^3$ is generated in a V2I environment comprising six vehicles and five BSs, which form 7 communication links. The carrier frequency is 28~GHz, and each endpoint uses a $4\times4$ uniform planar array (UPA) and a seven-beam codebook. Each sample pairs dual-endpoint RGB images with the corresponding $7\times7$ BPPM computed via~\eqref{eq:bppm}. Finally, DuSen-BP$^3$ comprises 6,965 valid samples, which are split into training, validation, and test sets in a 7:1:2 ratio.

RGB inputs are resized to $256\times448$. The four-stage backbone has depths $(2,2,8,2)$, where the first three stages use CIRA with routing top-$k$ values $(4,8,12)$, and the fourth adopts self-attention to refine the features. CiFormer is trained with AdamW for 150 epochs using a batch size of 64, a learning rate of $10^{-4}$, and weight decay of $10^{-3}$. The learning rate uses a five-epoch linear warm-up followed by cosine decay. For comparison, three representative vision backbones, i.e., Vision Transformer (ViT), MobileNetV4, and MambaOut, are adopted as baselines. BPPM prediction is evaluated using root mean squared error (RMSE) and mean absolute error (MAE) in dB. For the optimal beam-pair index $b_s^\star=\arg\max_{q,p}Y_{s,q,p}$ and the predicted top-$k$ beam-pair index set $\mathcal{T}_k(\widehat{\mathbf{Y}}_s)$ of sample $s$, beam-pair selection is evaluated using top-$k$ accuracy (Acc@$k$, \%) and beamforming (BF) loss (dB), defined as
\begin{equation}
\begin{aligned}
\mathrm{Acc}@k
&=\frac{100}{S}\sum_{s=1}^{S}
\mathbf{1}\{b_s^\star\in\mathcal{T}_k(\widehat{\mathbf{Y}}_s)\},\\
L_{\mathrm{BF},s}
&=\max_{q,p}Y_{s,q,p}-Y_{s,\hat q_s,\hat p_s}
\end{aligned}
\label{eq:evaluation_metrics}
\end{equation}
where $S$ is the number of evaluation samples and
$\mathbf{1}\{\cdot\}$ is an indicator.
Here we report Acc@1, Acc@3, and mean BF loss.

\subsection{Simulation Results}
% Reusable IEEEtran table (single column).
% Usage in the main document: \input{Tables/comparison_results}
% No additional packages are required. This is an input fragment, not a full document.
% Values are transcribed from the supplied comparison figure.
% Omitted Acc@5 (%) values: ViT 96.98; MobileNetV4 96.55; MambaOut 97.99; CiFormer 98.06.
\begin{table}[!t]
    \caption{Comparison of BPPM Prediction Performance}
    \label{tab:bppm_comparison}
    \centering
    \footnotesize
    \renewcommand{\arraystretch}{1.1}
    \setlength{\tabcolsep}{2pt}
    \newcommand{\comparisonheader}[1]{\begin{tabular}[c]{@{}c@{}}#1\end{tabular}}
    \begin{tabular*}{\columnwidth}{@{\extracolsep{\fill}}lcccccc@{}}
        \hline
        \comparisonheader{Model} & \comparisonheader{Params.\\(M)} & \comparisonheader{RMSE\\(dB)} & \comparisonheader{MAE\\(dB)} & \comparisonheader{Acc@1\\(\%)} & \comparisonheader{Acc@3\\(\%)} & \comparisonheader{Mean BF loss\\(dB)} \\
        \hline
        ViT & 10.995 & 6.318 & 4.143 & 68.25 & 91.81 & 0.943 \\
        MobileNetV4 & 14.950 & 6.553 & 4.156 & 68.39 & 91.16 & 0.924 \\
        MambaOut & 11.831 & 6.014 & 3.864 & 69.26 & 93.03 & 0.658 \\
        \textbf{CiFormer} & \textbf{10.986} & \textbf{5.810} & \textbf{3.638} & \textbf{72.41} & \textbf{94.68} & \textbf{0.641} \\
        \hline
    \end{tabular*}
    %\parbox{\columnwidth}{\footnotesize
        %Params.: number of model parameters; M: million; ViT: Vision Transformer; BF: beamforming. Bold indicates the best result in each column.
    %}
\end{table}

\begin{figure*}[!t]
	\centering	\includegraphics[width=1.0\textwidth]{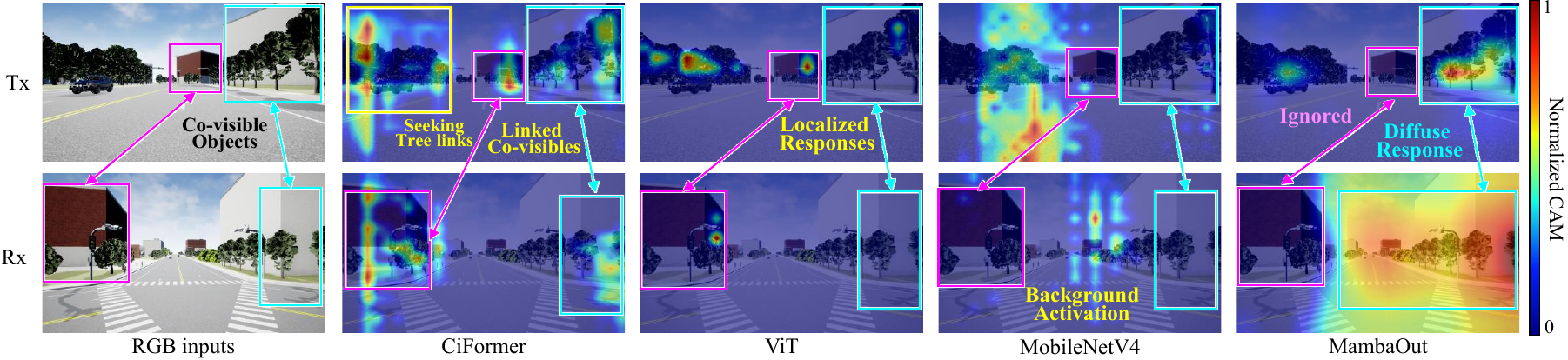}
	\caption{Grad-CAM visualizations of CiFormer, ViT, MobileNetV4, and MambaOut. CiFormer's activation maps emphasize the building and roadside trees in both views, whereas those of the baselines exhibit weaker cross-view correspondence.}
	\label{fig:GradCAM}
\end{figure*}

Table~\ref{tab:bppm_comparison} compares CiFormer with the three baseline backbones on DuSen-BP$^3$. It can be seen that CiFormer achieves lower RMSE, MAE, and mean BF loss and higher Acc@1 and Acc@3 than all three baselines. Relative to MambaOut, the strongest baseline, RMSE and MAE decrease by 3.39\% and 5.85\%, respectively. CiFormer's Acc@1 and Acc@3 reach 72.41\% and 94.68\%, exceeding MambaOut by 3.15 and 1.65 percentage points, respectively. CiFormer achieves these gains with a parameter count comparable to ViT and lower than the other baselines. ViT's global self-attention, MobileNetV4's spatial and channel mixing, and MambaOut's gated convolutions extract intra-view context, leaving cross-endpoint association to later feature fusion. CIRA instead uses opposite-view information to guide feature selection and enhancement, better aligning visual representations with the coupled Tx--Rx structure of beam-pair powers.

%This interpretation is further supported by the Grad-CAM~\cite{selvaraju2017gradcam} visualizations in Fig.~\ref{fig:GradCAM}. CiFormer highlights corresponding co-visible objects in both the Tx and Rx images, with more consistent responses across the two views than the baselines. These paired activations suggest that CIRA encourages both endpoint representations to focus on shared scene content, providing qualitative support for its use of opposite-view information to guide feature selection and enhancement.

This interpretation is further supported by the Grad-CAM~\cite{selvaraju2017gradcam} visualizations in Fig.~\ref{fig:GradCAM}. CiFormer highlights co-visible buildings and roadside tree clusters in both endpoint views, with more consistent activations than the baselines despite differences in image position and scale. ViT appears to rely on a few localized visual cues, as indicated by its isolated hotspots. MobileNetV4 also activates sky and road regions, suggesting distraction by backgrounds. MambaOut exhibits a broad, diffuse response in the Rx view. These paired activations suggest that CIRA encourages both endpoint representations to focus on shared scene content, providing qualitative support for its use of opposite-view information to guide feature selection and enhancement.

% \input{Tables/ablation_results}
% Single-column transposed alternative; both layouts are included for comparison.

% Transposed, single-column alternative to Tables/ablation_results.tex.
% Usage in the main document: \input{Tables/ablation_results_transposed}
% A distinct label allows both layouts to be included for comparison.
% No additional packages are required.
\begin{table}[!t]
    \caption{Ablation Study of Dual-Endpoint Inputs and CIRA Components}
    \label{tab:ciformer_ablation_transposed}
    \centering
    \footnotesize
    \renewcommand{\arraystretch}{1.1}
    \setlength{\tabcolsep}{2pt}
    \newcommand{\ablationheader}[1]{\begin{tabular}[c]{@{}c@{}}#1\end{tabular}}
    \begin{tabular*}{\columnwidth}{@{\extracolsep{\fill}}lccccc@{}}
        \hline
        \ablationheader{Variant} & \ablationheader{RMSE\\(dB)} & \ablationheader{MAE\\(dB)} & \ablationheader{Acc@1\\(\%)} & \ablationheader{Acc@3\\(\%)} & \ablationheader{Mean BF loss\\(dB)} \\
        \hline
        w/o Tx & 10.097 & 7.047 & 40.95 & 65.30 & 6.590 \\
        w/o Rx & 12.344 & 6.867 & 54.96 & 73.56 & 4.414 \\
        w/o Interaction & 6.001 & 3.740 & 70.98 & 93.89 & 0.723 \\
        \mbox{w/o top-$k$ routing} & 5.922 & 3.672 & 71.19 & 93.97 & 0.703 \\
        w/o Gating & 5.827 & 3.678 & 71.19 & 93.53 & 0.756 \\
        w/o SAB & 5.893 & 3.767 & 71.05 & 93.53 & 0.708 \\
        \textbf{Full CiFormer} & \textbf{5.810} & \textbf{3.638} & \textbf{72.41} & \textbf{94.68} & \textbf{0.641} \\
        \hline
    \end{tabular*}
    %\par\smallskip
    %\parbox{\columnwidth}{\footnotesize
        %Full CiFormer: complete model; w/o: without; SAB: semantic-aware bias; BF: beamforming. Bold indicates the best result in each column.
    %}
\end{table}

To further validate the effectiveness of CiFormer, an ablation study is also conducted. As shown in Table~\ref{tab:ciformer_ablation_transposed}, using only one endpoint image substantially degrades CiFormer's prediction and beam-selection performance. This degradation in both single-endpoint settings highlights the necessity of dual-endpoint sensing. In addition, RMSE increases more when the Rx view is removed, indicating greater sensitivity to the loss of receiver-side information. Disabling cross-view interaction degrades prediction and beam-selection performance, highlighting the importance of associating features across endpoints. Removing top-$k$ routing turns CIRA into dense cross-view attention over all windows and reduces performance, suggesting that irrelevant background regions can interfere with propagation-relevant feature association. The degradation without gating or SAB supports their respective roles in suppressing unreliable queries and modeling semantic correspondence across viewpoints without a shared geometric reference. These results demonstrate the effectiveness of CiFormer's link-centric design, which combines dual-endpoint observations with selective cross-view association to improve BPPM prediction and beam-pair selection, suggesting its applicability to broader sensing-aided communication tasks.

\begingroup
\makeatletter

\def\section{\@startsection{section}{1}{\z@}
  {1.5ex}
  {0.5ex}
  {\normalfont\normalsize\centering\scshape}}

\patchcmd{\thebibliography}
  {0.3\baselineskip plus 0.1\baselineskip minus 0.1\baselineskip}
  {0.1\baselineskip}
  {}{}

\makeatother

\section{Conclusion}
In this letter, we proposed CIRA for selective association and enhancement of propagation-relevant features across dual-endpoint views. CiFormer applies CIRA to link-centric V2I BPPM prediction from paired RGB images. Experiments on DuSen-BP$^3$ showed improved prediction and beam selection with fewer parameters. Grad-CAM and ablation results further supported dual-endpoint sensing and selective interaction. These results demonstrate the effectiveness of CIRA's design, suggesting its applicability to broader wireless  communication tasks that rely on environment sensing.


\bibliographystyle{IEEEtran}
\bibliography{ref}

\endgroup

\vfill

\end{document}